\documentclass[10pt,twocolumn,english,preprintnumbers,amsmath,amssymb,prb]{revtex4}
\usepackage[T1]{fontenc}
\usepackage[latin9]{inputenc}
\usepackage{color}
\usepackage{babel}

\usepackage{amstext}
\usepackage{graphicx}
\usepackage[unicode=true, pdfusetitle,
 bookmarks=true,bookmarksnumbered=false,bookmarksopen=false,
 breaklinks=false,pdfborder={0 0 1},backref=false,colorlinks=true]
 {hyperref}
\hypersetup{
 citecolor=blue}

\makeatletter
\@ifundefined{textcolor}{}
{%
 \definecolor{BLACK}{gray}{0}
 \definecolor{WHITE}{gray}{1}
 \definecolor{RED}{rgb}{1,0,0}
 \definecolor{GREEN}{rgb}{0,1,0}
 \definecolor{BLUE}{rgb}{0,0,1}
 \definecolor{CYAN}{cmyk}{1,0,0,0}
 \definecolor{MAGENTA}{cmyk}{0,1,0,0}
 \definecolor{YELLOW}{cmyk}{0,0,1,0}
 }

\usepackage{graphicx}
\usepackage{dcolumn}
\usepackage{bm}
\usepackage{wasysym}
\usepackage{textcomp}
\usepackage{epstopdf}
\usepackage{stmaryrd}
\usepackage{txfonts}
\preprint{APS/123-QED}

\makeatother

\begin{document}

\title{Observation of tunable exchange bias in Sr$_{\text{2}}$YbRuO$_{\text{6}}$}

\author{R.\,P. Singh and C.\,V. Tomy}

\affiliation{Department of Physics, Indian Institute of Technology Bombay, Mumbai-400076,
India.}

\author{A.\,K. Grover}

\affiliation{Department of Condensed Matter Physics and Materials Science, Tata
Institute of Fundamental Research Colaba, Mumbai-400005, India.}
\begin{abstract}
The double perovskite compound, Sr$_{2}$YbRuO$_{6}$, displays reversal
in the orientation of magnetic moments along with negative magnetization
due to an underlying magnetic compensation phenomenon. The exchange
bias (EB) field below the compensation temperature could be the usual
negative or the positive depending on the initial cooling field. This
EB attribute has the potential of getting tuned in a preselected manner,
as the positive EB field is seen to crossover from positive to negative
value above $T_{\mathrm{comp}}$.
\end{abstract}
\maketitle
The notion of an exchange bias (EB) field \cite{Nogues J3M}, which
is exemplified by the identification of a shift \cite{Meiklejohn}
in the centre of gravity of the magnetization hysteresis ($M$-$H$)
loop, has implications for magnetic read heads \cite{read heads},
thermally assisted magnetic random access memories \cite{TARAM},
and other spintronics devices \cite{Spin Valve1,Spin Valve2}. The
EB phenomenon is considered to be a quintessential attribute \cite{Nogues J3M}
of a ferromagnetic (FM)/ antiferromagnetic (AFM) composite or a bilayer
assembly, when the entire system is cooled below ordering temperature
$T_{N}$ of the AFM part, which lies below the ($T_{C}$) of the FM,
i.e., $T_{N}<T_{C}$. Usually, the EB field is negative as the centre
of $M$-$H$ loop gets left shifted below ($T_{N}$) \cite{Nogues J3M}.
But occasionally the right shift of the loop, i.e., positive exchange
bias (PEB) field has also been noted \cite{Nogus,P_Leighton,P_Gredig,P_Radu,P_Ali,P_Kohlhepp}.
The EB had mostly been observed in magnetic multi-component systems,
such as oxidized magnetic nanoparticles, FM/AFM multilayers, FM/spin
glass bilayer thin films, etc. \cite{Nogues J3M}. However interestingly,
in recent years, it has also been noted in several bulk homogenous
materials, like, manganites \cite{Mangnite}, cobaltates \cite{cobaltate}
and admixed intermetallic compounds \cite{Chen,Kulkarni ieee}. Recently
Kulkarni \emph{et\,al}.~\cite{Kulkarni} have unearthed EB effect
in a single crystal of an admixed rare earth intermetallic compound,
Nd$_{0.75}$Ho$_{0.25}$Al$_{2}$, in close proximity to the compensation
temperature ($T_{\mathrm{comp}}$) in it. They encountered the sudden
surfacing of the PEB on approaching $T_{\mathrm{comp}}$, the EB changed
sign on going across it; the sign reversal of exchange bias field
is sought to be rationalized in terms of reversal in the\textcolor{black}{{}
nearly balanced} local magnetic moment contributions from antiferromagnetically
linked Nd$^{3+}$ and Ho$^{3+}$ ions along with that of the conduction
electron polarization (CEP), with respect to the
applied field direction \cite{Kulkarni}.

We have now studied the EB phenomenon in a double perovskite antiferromagnetic
(AFM) insulator Sr$_{\text{2}}$YbRuO$_{\text{6}}$ \cite{CVT JPCM}.
This compound belongs to the Ruthenates family, Sr$_{2}$\emph{Ln}RuO$_{6}$
(\emph{Ln} = Y or rare earth) \cite{Donohue,DAE,PhD}, and shows interesting
magnetic properties below the magnetic ordering temperature, which
include the (i) magnetization reversal and (ii) the magnetic \textcolor{black}{compensation
characterized by crossover of the magnetization axis ($M=0$ line)
towards the negative magnetization value}s \cite{CVT JPCM}. In the
antiferromagnetic Sr$_{\text{2}}$YRuO$_{\text{6}}$ compound, containing
non-magnetic Y element, the local moment resides on Ru ion alone \cite{Battle},
whereas in Sr$_{\text{2}}$YbRuO$_{\text{6}}$, both Yb and Ru ions
possess local magnetic moments of $4.54\,\mu_{B}$ and $3.87\,\mu_{B}$,
respectively \cite{Doi }. The monoclinic structure \cite{Doi } of
these AFM insulators facilitates the Dzyaloshinsky--Moria (D--M) interaction
\cite{Tung} between the antiferromagnetically ordered local moments.
The anisotropic D--M interaction results in canting of the antiferromagnetically
ordered moments and hence a weak residual ferromagnetism germinates
\cite{Dzyaloshinsky,Moriya}. The presence of a weak ferromagnetic
component could provide a circumstance analogous to that associated
with the FM/AFM composite systems. Though the magnetization reversal,
zero magnetization and negative magnetization in Sr$_{\text{2}}$YbRuO$_{\text{6}}$
could be expected from the combined effects of D--M interaction and
the unidirectional anisotropy \cite{CVT JPCM}, the discovery of the
EB and its sign reversal near the compensation temperature and the
dependence of sign of EB at low temperature (i.e., below the $T_{\mathrm{comp}}$)
on the thermo-magnetic history has brought to light the possibility
of tuning of the EB from positive to negative values in a convenient
and predictable manner in double perovskite compounds. Such an interesting
finding is being reported here; this attribute could be utilized in
niche applications in spintronics.

\begin{figure}
\includegraphics[scale=0.3]{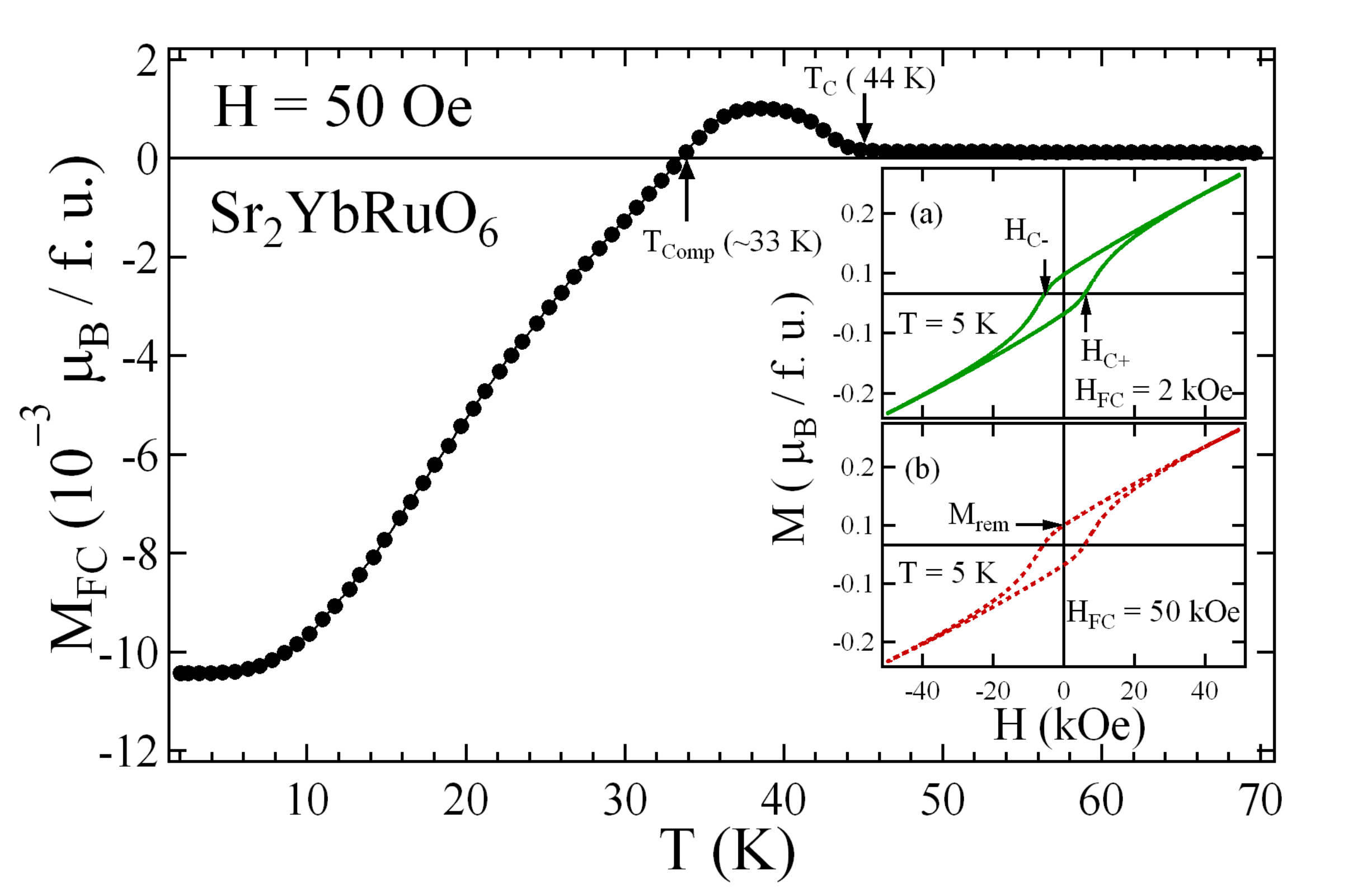}

\caption{\label{fig:1}Temperature variation of the field-cooled (FC) magnetization
in $H=50$\,Oe in Sr$_{\text{2}}$YbRuO$_{\text{6}}$. The $T_{C}$
and $T_{\mathrm{comp}}$ values have been identified. The insets (a)
and (b) show the magnetic hysteresis ($M$-$H$) loops traced at 5\,K
after initial field cooling in (a) $H_{\mathrm{FC}}=2$\,kOe and
(b) 50\,kOe, respectively.}

\end{figure}

A typical field-cooled (FC) magnetization curve $M_{\mathrm{FC}}$
measured at low fields (e.g. at $H$\,=\,50\,Oe) in Sr$_{\text{2}}$YbRuO$_{\text{6}}$
is shown in Fig.~\ref{fig:1}. The sudden increase in the $M_{\mathrm{FC}}$
at the onset of the ordering temperature (44\,K) is due to the presence
of a weak ferromagnetic component \cite{CVT JPCM} in this compound.
The start of decrease in magnetization around $T\sim39$\,K is due
to the onset of reversal in magnetization of the residual ferromagnetic
part. This\textcolor{red}{{} }\textcolor{black}{is triggered} by the
D--M interaction, as noted earlier in the case of isostructural Sr$_{\text{2}}$YRuO$_{\text{6}}$
compound \cite{CVT_PRB}. On cooling below 39\,K, the decrease in
magnetization continues and it crosses the $M=0$ axis at the compensation
temperature of $\sim\,33.5$\,K (marked as $T_{\mathrm{comp}}$),
and magnetization values move towards negative values on further lowering
the temperature, signalling the importance of the role of Yb moments
in Sr$_{\text{2}}$YbRuO$_{\text{6}}$; no crossover of $M=0$ axis
occurs in Sr$_{\text{2}}$YRuO$_{\text{6}}$ \cite{CVT_PRB}, where
Y$^{3+}$ ions are non-magnetic. %

\begin{figure}[h]
\includegraphics[scale=0.3]{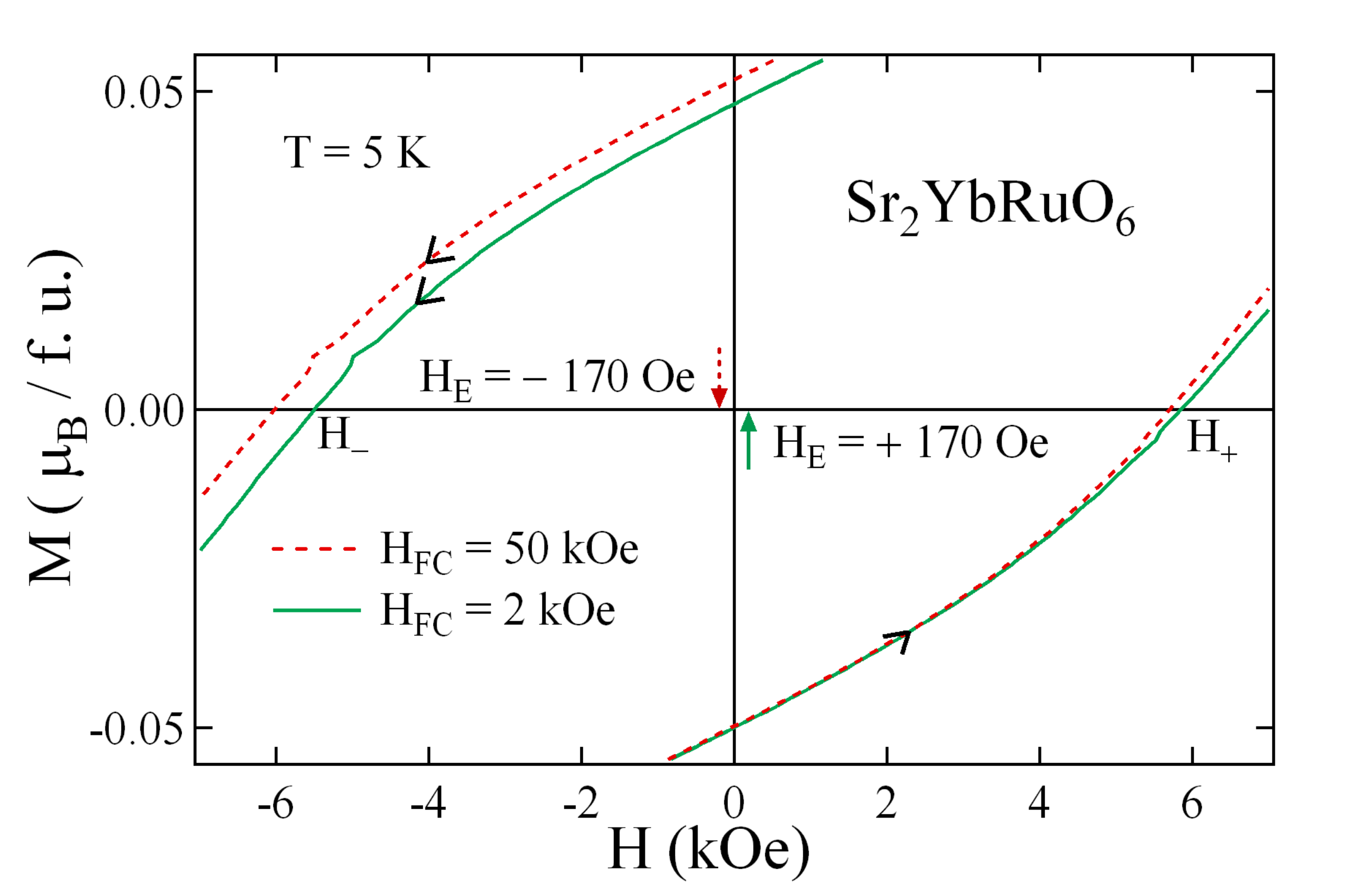}

\caption{\label{fig:2}Expanded portion of the magnetization loops in cooling
fields ($H_{\mathrm{FC}}$) of 2\,kOe and 50\,kOe in Sr$_{\text{2}}$YbRuO$_{\text{6}}$
at 5\,K. Note that the exchange bias field $H_{E}$ in the two \textcolor{black}{cases}
have opposite signs.}

\end{figure}

The two insets in Fig.~\ref{fig:1} show the $M$-$H$ \textcolor{black}{loops
recorded between $\pm\,50$\,kOe} at 5\,K in Sr$_{\text{2}}$YbRuO$_{\text{6}}$
for two typical initial cooling fields, viz., 2\,kOe and 50\,kOe.\textcolor{black}{{}
At very high fields ($H>20$\,kOe), the} $M$-$H$\textcolor{black}{{}
response in both the insets of Fig.~\ref{fig:1} are quasi-linear,
as in antiferromagnets. However, at lower fields (i.e., between $\pm\,20$\,kOe),
one can notice the presence of hysteresis bubbles, with small remnant
magnetization ($~\,0.05\,\mu_{B}$/f.u. of }Sr$_{\text{2}}$YbRuO$_{\text{6}}$\textcolor{black}{),
which can be attributed to the residual ferromagnetic component originating
from a slight canting of the antiferromagnetically coupled large moments
of Yb and Ru ions. There is a subtle but important difference between
the two bubbles in the two insets of Fig.~\ref{fig:1}, which pertains
to the shift in the centre of gravity of these bubbles. Figure~\ref{fig:2}
shows the central portions of two hysteresis bubbles plotted together
on an expanded scale, a noticeable difference between the dashed and
solid curves as they cross the $M=0$ axis at a magnetic field ($H_{-}$
or $H_{+}$) can be immediately recognized. The difference between
the two loops in Fig.~\ref{fig:2} can be characterized in terms
of an exchange bias field ($H_{E}$), which is usually defined as
$H_{E}=(H_{-}+H_{+})/2$, where $H_{-}$ and $H_{+}$ are the field
values at which magnetization axis is crossed during the descending
and the ascending field cycles.}\textcolor{red}{{} }The loop measured
between $\pm\,50$\,kOe in a cooling field ($H_{\mathrm{FC}}$) of
2\,kOe shows that the offset field \textcolor{red}{$H_{E}$} is positive
and the $M$-$H$ loops is slightly right shifted, whereas, in that
with $H_{\mathrm{FC}}$ = 50\,kOe the offset field,\textcolor{red}{{}
}$H_{E}$, is negative. The two $H_{E}$ values (positive/negative
exchange bias) with two different initial FC values are marked in
Fig.~\ref{fig:2}.%

\begin{figure}
\includegraphics[scale=0.4]{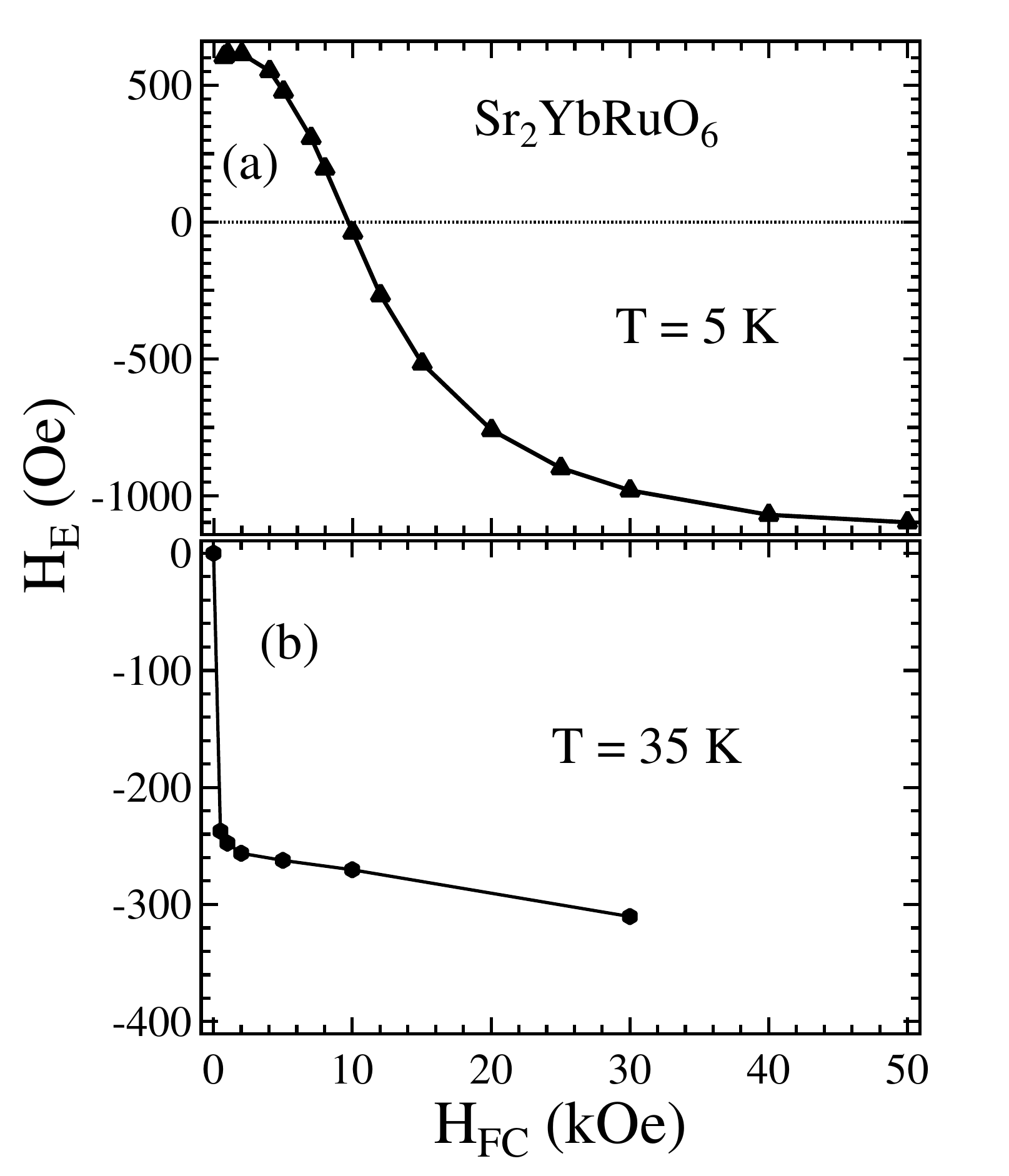}

\caption{\label{fig:3(a)}\label{fig:(3b)}Dependence of the exchange bias
field ($H_{E}$) on the initial cooling field ($H_{\mathrm{FC}}$)
in Sr$_{\text{2}}$YbRuO$_{\text{6}}$ at (a) 5\,K and (b) 35\,K.}

\end{figure}

To explore the effect of $H_{\mathrm{FC}}$ on $H_{E}$, $M$-$H$
loops were later traced between $\pm\,15$\,kOe as a function of
$H_{\mathrm{FC}}$ at two temperatures, 5\,K (i.e., well below the
compensation temperature $T_{\mathrm{comp}}$ of 33.5\,K) and 35\,K
($>T_{\mathrm{comp}}$) as a function of the cooling field ($H_{\mathrm{FC}}$).
The $H_{E}$ values obtained from such measurements are shown in Figs.~\ref{fig:3(a)}(a)
and \ref{fig:(3b)}(b). At 5\,K (\emph{cf.} Fig~\ref{fig:3(a)}(a)),
$H_{E}$ is positive for low cooling fields. As the value of $H_{\mathrm{FC}}$
increases, $H_{E}$ monotonically\textcolor{black}{{} decreases, crossing
zero at }$H_{\text{FC}}\sim10$\,kOe, and thereafter moves towards
a saturated negative value, above $H_{{\rm FC}}$ of 30\,kOe. On
the other hand, at $T=35$\,K, one observes only the usual negative
$H_{E}$ values in low as well as high cooling fields ($cf.$ Fig.~\ref{fig:(3b)}(b)).
We also noted that a given \textcolor{black}{$H_{E}$ value depends
on the extent of the limit set on the cyclic field (e.g. $\pm15$\,kOe
for data in Fig.~\ref{fig:3(a)}(a)), however the trend of change
in sign of $H_{E}$ as a function of }$H_{\mathrm{FC}}$ (at 5\,K)
and its approach to saturation (above $T_{\mathrm{comp}}$ at 35\,K)
does not depend on the choice of the limit set on the cycling field.%

\begin{figure}
\includegraphics[scale=0.45]{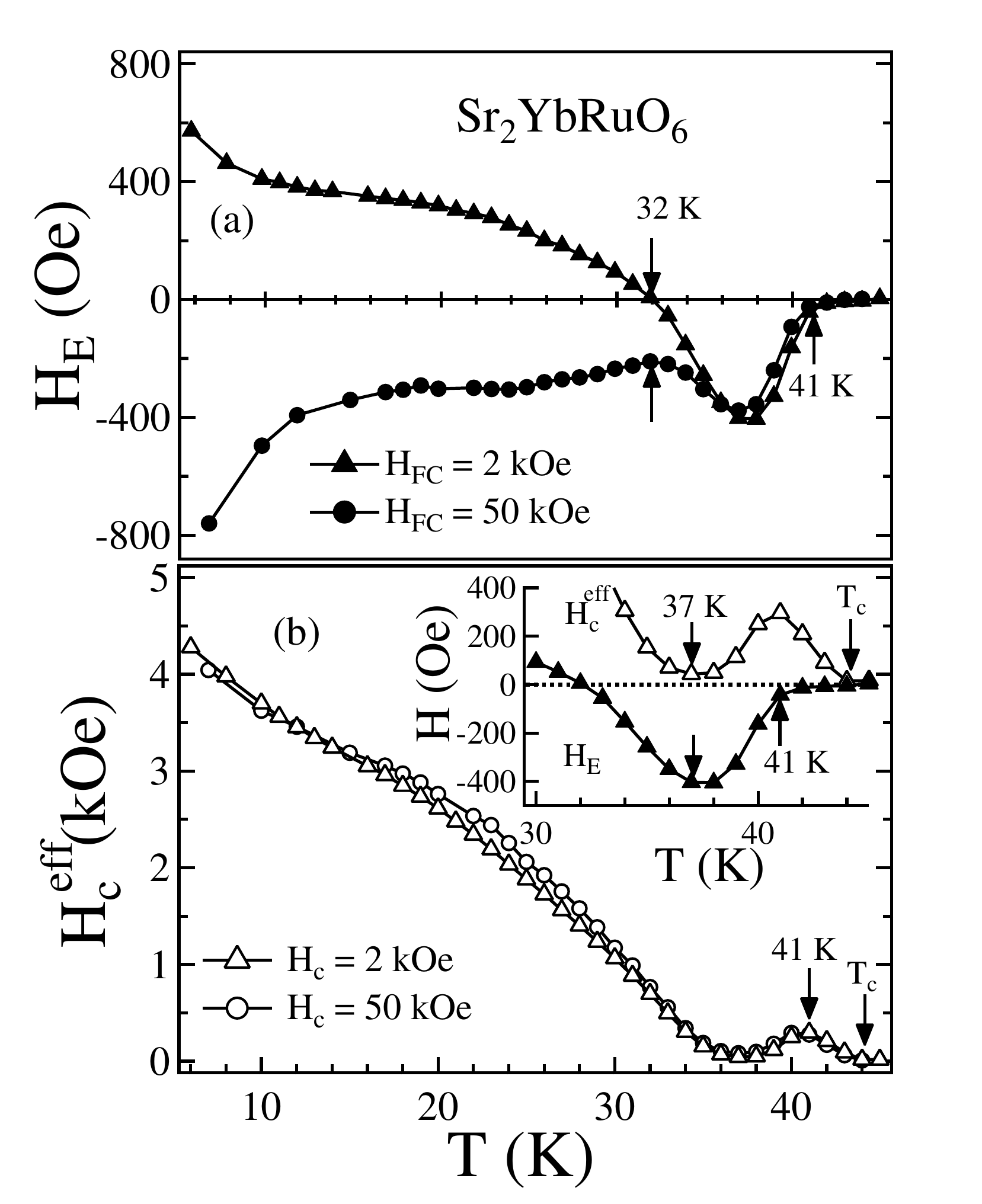}\caption{\label{fig:(4a)}\label{fig:(4b)}Temperature variation of (a) the
exchange bias field ($H_{E}$) and (b) the effective coercive field
($H_{c}^{\mathrm{eff}}$) in cooling fields ($H_{\mathrm{FC}}$) of
50\,kOe and 2\,kOe. The inset panel in Fig.~\ref{fig:(4b)}(b)
shows $H_{E}$ (closed symbols) and $H_{c}^{\mathrm{eff}}$ (open
symbols) in the temperature range 30--45\,K for $H_{\mathrm{FC}}=2$kOe}

\end{figure}

The temperature variation study of the EB effect was performed on
the hysteresis loops measured between $\pm\,15$\,kOe in two cooling
fields, 50\,kOe and 2\,kOe. $H_{E}$ values obtained from the hysteresis
loops at different temperatures are shown in Fig.~\ref{fig:(4a)}(a)
along with the half width of the hysteresis loops, which is \textcolor{black}{often}
termed as the effective coercive field, $H_{c}^{\mathrm{eff}}=-(H_{-}-H_{+})/2$,
in Fig~\ref{fig:(4b)}(b) for comparison \cite{Webb}. We note first
in Fig.~\ref{fig:(4b)}(b) that just as in the two insets of Fig.~\ref{fig:1}
for 5\,K, the width of hysteresis loop does not imbibe any dependence
on the initial cooling field values at all temperatures. $H_{c}^{\mathrm{eff}}(T)$
in Fig.~\ref{fig:(4b)}(b) displays a minimum at about 37\,K (see,
also, inset of Fig.~\ref{fig:(4b)}(b)), which is unexpectedly above
the compensation temperature of 33\,K in low fields (\emph{cf.} Fig.~\ref{fig:1})
\cite{CVT JPCM}. $T_{\mathrm{comp}}$ values had earlier been noted
\cite{CVT JPCM} to decrease with large enhancement in $H$ ($\mbox{\ensuremath{H}>2\,\ kOe}$)
in this system. As\emph{ }regards the temperature dependence of EB
field in Fig.~\ref{fig:(4a)}(a), the \textcolor{red}{$H_{E}$ }values
for $H_{\mathrm{FC}}=50$\,kOe maintained the usual negative sign
over entire temperature range on warming up from 5\,K towards $T_{C}$.
However, those for $H_{\mathrm{FC}}=2$\,kOe show a sign reversal
from positive to negative values at about 32\,K, which is close to
the $T_{\mathrm{comp}}$ value in Fig.~\ref{fig:1}. The latter behaviour
appears to echo the change in sign of \textcolor{red}{$H_{E}$}\textcolor{black}{{}
reported across }$T_{\mathrm{comp}}$ in the admixed rare earth intermetallic
Nd$_{0.75}$Ho$_{0.25}$Al$_{2}$ \cite{Kulkarni} and other similar
systems which imbibe compensation behaviour \cite{Kulkarni ieee}.
However, the sense of sign change in Sr$_{\text{2}}$YbRuO$_{\text{6}}$
is phase reversed \emph{vis.\,a\,vis. }that in Nd$_{0.75}$Ho$_{0.25}$Al$_{2}$.
The $\left|H_{E}\right|$ values for $H_{\mathrm{FC}}=50$\,kOe reach
a local minimum near $T_{\mathrm{comp}}$ ($cf.$ Fig.~\ref{fig:(4a)}(a)).
Above about 35\,K, the two sets of $H_{E}(T)$ values (in Fig.~\ref{fig:(4a)}(a))
overlap, thereby implying their independence of the initial field
cool history. The $\left|H_{E}(T)\right|$ values can be seen to reach
a local maximum at 37\,K, where $H_{c}^{\mathrm{eff}}(T)$ is minimum.
This correspondence between $H_{c}^{\mathrm{eff}}(T)$ and $\left|H_{E}(T)\right|$
is also very different from the trend seen in the admixed rare-earth
intermetallic and the magnetic multilayer systems \cite{Kulkarni,Nogus,Kulkarni ieee}
across the respective $T_{\mathrm{comp}}$ values. In the latter varieties
of systems \cite{Kulkarni,Kulkarni ieee,Webb}, the width of the hysteresis
loop nearly collapses at $T_{\mathrm{comp}}$, and the exchange bias
also tends to disappear.%

In consonance with the vanishing of the magnetization at the compensation
temperature, Fig.~\ref{fig:(4a)}(a) also shows that while cooling
down $\left|H_{E}(T)\right|$ surfaces up only on cooling field below
41~K, the temperature at which $H_{c}^{\mathrm{eff}}(T)$ shows a
local maximum. It has been argued earlier \cite{CVT JPCM} that the
decrease in magnetization below 41\,K signals the onset of turnaround
in the orientation of local moments of Ru due to coming in the play
of the D--M interaction. We believe that subtle differences between
the details of correspondence between the $H_{c}^{\mathrm{eff}}(T)$
and $H_{E}(T)$ occurring in Sr$_{\text{2}}$YbRuO$_{\text{6}}$ \emph{vis.
a vis.} other systems, where compensation behavior has been noted,
are a consequence of competition and interplay between three components
in Sr$_{\text{2}}$YbRuO$_{\text{6}}$ (\emph{viz. }$(i)$ residual
ferromagnetism from Ru and $(ii)$ residual magnetisation from Yb
moments, and $(iii)$ reorientation in Ru and Yb moments triggered
by the D--M interaction) instead of the usual competition between
the two antiferromagnetically linked components in other systems \cite{Ren}
.

To summarize, we have shown that an antiferromagnetic double perovskite
compound Sr$_{\text{2}}$YbRuO$_{\text{6}}$ comprising magnetic moments
residing on rare earth and transition elements exhibits a large variety
in exchange bias behaviour. The dependence of EB field on the cycling
field, its sign change on moving across $T_{\mathrm{comp}}$, dependence
of EB on the cooling field, etc. are attributes, which have been noted
earlier in different magnetic multi-components systems, in the forms
of multilayer, coated nano particles, etc. \cite{Nogues J3M} as well
as in single component systems, like, admixed rare earth intermetallics
\cite{Kulkarni,Kulkarni ieee}, transition metal oxides \cite{cobaltate,DAE,Mangnite},
etc. Our above findings that an antiferromagnetic double perovskite
compound Sr$_{\text{2}}$YbRuO$_{\text{6}}$, exhibits the entire
spectrum of exchange bias behaviour in the same compound, underscores
the importance and novelty of oxide compounds comprising rare earth
and transition metal ions. The exchange bias field of either sign
can be selected in the Sr$_{\text{2}}$YbRuO$_{\text{6}}$ compound
in a predetermined manner by choosing the cooling field and/or changing
the temperature. The attribute of easy tuning of the exchange bias
calls for further explorations of this behaviour in single crystal
and or/oriented thin film form to gain deeper insight into the microscopic
basis of the origin of exchange bias and its variation with field/temperature
and thermomagnetic history in the compounds crystallizing in double
perovskite structure.

\end{document}